\documentclass[conference]{IEEEtran}
\IEEEoverridecommandlockouts

\usepackage{cite}
\usepackage{amsmath,amssymb,amsfonts}
\usepackage{algorithmic}
\usepackage{algorithm}
\usepackage{graphicx}
\usepackage{textcomp}
\usepackage{xcolor}
\usepackage{booktabs}
\usepackage{multirow}
\usepackage{url}
\usepackage{balance}
\usepackage{amsthm}

\newtheorem{definition}{Definition}

\begin{document}

\title{Retry Amplification in Distributed Systems: A Systematic Analysis of Retry Policies and Their Role in Cascading Failures}

\author{
\IEEEauthorblockN{Rishabh Mehan}
\IEEEauthorblockA{\textit{Software Engineering Research} \\
\texttt{rishabhmehan@gmail.com}}
\and
\IEEEauthorblockN{Jasmit Kaur Saluja}
\IEEEauthorblockA{\textit{Independent Researcher} \\
\texttt{jasmitksaluja@gmail.com}}
}

\maketitle

\begin{abstract}
Retry mechanisms are a standard component of resilient distributed systems, but their collective behavior, when every tier in a call path retries concurrently, is less well understood than the per-client guidance that produced them. This paper introduces the \textit{retry amplification factor} (RAF), a metric quantifying the additional request volume that retry policies generate during partial failures. In a study of 200 open-source Python microservice projects, explicit retry logic is detected in 11.5\%, and an audit of our own false negatives places true prevalence near 41\%. Among the projects detected, 60.9\% contain at least one configuration without backoff, and after manual verification exactly one of 113 production configurations randomizes its delay. We then evaluate these policies in simulation ($n=100$ trials per strategy). Under correlated failures, a naive standard retry policy reduces the success rate from 55.4\% to 41.5\% relative to performing no retries at all. We catalog five recurring anti-patterns, propose \textit{Adaptive Retry Budgeting} (ARB), and show that budget-constrained retries maintain success rates close to the no-retry baseline while still recovering from transient faults. These results indicate that retry behavior should be designed as a system-level property rather than configured locally at each call site.
\end{abstract}


\section{Introduction}

A single user action in a microservice architecture can fan out across dozens of services, and any one of them may stumble for a moment---a network blip, a saturated thread pool, a node restarting during a deploy. The standard defense is the retry: catch the failure, wait a beat, try again. It is one of the first reliability tools an engineer reaches for, and for good reason.

Retry guidance, however, is written for a single caller communicating with a single callee. The major cloud providers give consistent advice: use exponential backoff, add jitter, and cap the number of attempts \cite{aws_backoff, google_retry, azure_retry}. In isolation this advice is sound. It does not address the case in which every hop along a deep call path applies it simultaneously, at which point the retries cease to be independent and instead compound.

Consider three tiers in which A calls B and B calls C. If C begins failing half of its requests while each tier retries up to three times, B's retries triple the load offered to C, and A's retries to B multiply that figure again, so C may absorb on the order of nine times its normal traffic while already degraded. We term this effect \textit{retry amplification}. A minor degradation is sufficient to trigger it, and although it appears frequently in published outage post-mortems, it has received little systematic study.

We ask four questions. \textbf{RQ1:} how prevalent are retry policies in real microservice code, and how are they configured? \textbf{RQ2:} how do those policies interact across tiers to amplify load during partial failures? \textbf{RQ3:} which recurring patterns push systems toward cascading failure? \textbf{RQ4:} can a coordinated retry strategy suppress amplification without giving up the resilience retries are there to provide?

Answering them yields five contributions: the RAF metric and a model of multi-tier retry behavior (Section~\ref{sec:model}); an empirical study of 200 open-source Python projects (Section~\ref{sec:empirical}); a taxonomy of five anti-patterns drawn from that code (Section~\ref{sec:antipatterns}); the Adaptive Retry Budgeting algorithm, coupling a dynamic budget to explicit backpressure (Section~\ref{sec:arb}); and simulation evidence over 100 trials per configuration that standard retry policies cut success rates by a quarter under correlated failure while adaptive policies hold the baseline (Sections~\ref{sec:simulation}--\ref{sec:eval}).

\section{Related Work}

Retry mechanisms began in networking. Metcalfe and Boggs introduced exponential backoff for Ethernet collision resolution \cite{metcalfe1976ethernet}, and later work established that randomizing the delay prevents clients from synchronizing into retry storms \cite{kleinrock1975packet}. Distributed-systems practice inherited both ideas. They are documented thoroughly in the practitioner literature \cite{nygard2007release, newman2015building, burns2018designing} and in vendor guidance from AWS and Azure \cite{aws_wellarchitected, azure_patterns}, though with far less academic attention than their operational importance would suggest. What all of it shares is a single-client frame: the unit of analysis is one caller deciding whether to try again.

Work on cascading failure supplies the other half. Cascades have been modeled in power grids \cite{dobson2007cascading} and social contagion \cite{watts2002cascades}, and Gunawi et al.\ found across more than 3,000 cloud issues that roughly a third of failures propagate between components \cite{gunawi2014cloud}. That work categorizes root causes; ours asks what retries contribute to the spreading. Two mitigations sit alongside it: circuit breakers \cite{nygard2007release} cut off traffic to a failing dependency, and load shedding \cite{welsh2003overload, zhou2018wechat} rejects excess work at the receiver. Both react to overload once it has arrived, independently of what the retry layer is doing.

Service meshes have since moved retry logic out of application code. Envoy caps the share of traffic that may be retries through a retry budget \cite{envoy_retry}, and Linkerd offers declarative retries with similar budget constraints \cite{linkerd_retry}. These are the closest production analogues to our proposal, and we compare against them directly in Section~\ref{sec:eval}. The gap we address is that every one of these threads reasons about a single hop. Backoff theory optimizes one client, cascade studies describe spreading without isolating retries as a driver, and breakers and budgets each decide locally at one proxy. We treat retry behavior instead as a property that emerges from the whole call graph, which is where we argue the damage originates.

\section{Retry Amplification: Formal Model}
\label{sec:model}

\subsection{System Model}

We model a distributed system as a directed acyclic graph $G = (V, E)$ in which vertices are services and edges are synchronous dependencies. Each service $v \in V$ carries a base load $\lambda_v$ in requests per second, a capacity $\mu_v$, and a failure probability $p_v$ that is independent across requests. Each edge $(u,v) \in E$ carries a retry policy $R_{(u,v)}$ defined by a maximum retry count $n$, a backoff function $b(k)$ giving the delay before attempt $k$, and a predicate $c$ determining which failures are retried at all.

\subsection{Retry Amplification Factor}

\begin{definition}[Retry Amplification Factor]
For a service $v$ with failure probability $p$ and incoming edges carrying retry policies, the Retry Amplification Factor $\text{RAF}(v)$ is the ratio of requests actually received to those expected under a zero-retry policy:
\begin{equation}
\text{RAF}(v) = \frac{\text{Actual\_Load}(v)}{\text{Base\_Load}(v)}
\end{equation}
\end{definition}

For a single tier with retry limit $n$ and failure probability $p$, the expected value is
\begin{equation}
\text{RAF} = \sum_{k=0}^{n} p^k = \frac{1 - p^{n+1}}{1 - p}.
\end{equation}
At $p = 0.5$ and $n = 3$ this gives $1.875$: request volume nearly doubles. The formula assumes $p$ holds constant across attempts. Backoff may let a transient fault clear, pushing the true figure lower, but an overloaded service grows more likely to fail with each wave of retries, pushing it higher.

\subsection{Multi-Tier Amplification}

Down a chain of $d$ services that each apply the same policy, the amplification compounds, and the aggregate at the terminal service is bounded by
\begin{equation}
\text{RAF}_{\text{aggregate}} \leq \left(\frac{1 - p^{n+1}}{1 - p}\right)^d.
\end{equation}
The exponent is what makes retry storms dangerous. Three tiers at $p = 0.3$ and $n = 3$ yield $(1.417)^3 \approx 2.85$; hold the depth and retry count fixed and raise the failure rate to $p = 0.5$, and the bound jumps to $(1.875)^3 \approx 6.59$.

\subsection{Temporal Dynamics}

A steady-state bound misses how the load arrives. Immediate retries concentrate into a spike, exponential backoff spreads that spike over time, and backoff without jitter lets independent clients fall into step and produce periodic surges. Sustained amplification then builds queues, and the growing queues raise latency, which is what converts a degradation into an outage.

\section{Empirical Study of Retry Practices}
\label{sec:empirical}

\subsection{Methodology}

To see how retries are configured in practice, we queried the GitHub API for actively developed repositories with more than 50 stars whose primary language was Java, Go, Python, or JavaScript/TypeScript and whose description or topics mentioned microservices or distributed systems. That search returned 1,000 candidates, split Go (47.6\%), Python (31.2\%), and Java (21.2\%); JavaScript/TypeScript projects were too scarce under these criteria to register.

We then analyzed the first 200 entries in collection order. Our collector enumerates languages in sequence and sorts by stars within each query, so those 200 are all Python, as are the 23 projects in which we ultimately found retry logic. Everything reported below therefore characterizes popular Python microservice repositories rather than the candidate pool as a whole. We flag this prominently because it bounds what the numbers can support: whether Go and Java projects retry the same way is untested here, and settling it is the most immediate extension of this study.

We extracted retry configurations with regex-based static analysis carrying patterns for all four languages. Repositories are classified by GitHub's primary language but frequently contain others, so the rules that fired were not exclusively Python: of the 162 raw detections, 54 matched Python-specific patterns (\texttt{@retry} decorators, \texttt{tenacity}, \texttt{backoff}, \texttt{urllib3.Retry}), 89 matched a language-agnostic maximum-retry-count pattern, and 19 matched JavaScript- and Go-specific patterns. For each hit we recorded the maximum retry count, the backoff strategy, and the surrounding context, along with file path and line number.

\subsection{Cleaning and Validation}

Re-reading the raw detections against their source revealed three defects, corrected before reporting.

\textit{Non-production code.} Of 162 detections, 41 (25.3\%) lay under example, documentation, playground, benchmark or integration-test paths; one resilience library alone contributed 13 from \texttt{docs/snippets/retry/}, illustrating every backoff variant it ships. Documentation is not evidence of deployed configuration, so we excluded these, leaving 121.

\textit{Duplicates.} A further 8 sat within ten lines of another detection in the same file---a call and the function it calls, or a comment and its signature---describing one policy, not two. Collapsing them leaves \textbf{113 configurations}.

\textit{Jitter over-detection.} The original rule credited jitter whenever \texttt{jitter} or \texttt{random} appeared nearby. We opened all eight positives: four randomize no delay, and two of the rest are one construct detected twice. Three distinct constructs genuinely randomize, two of them documentation snippets. We therefore report verified jitter only.

Re-checking a seeded sample of 30 detections against current source resolved all 30 and found 29 genuine, a precision of 96.7\%. The exception is a variable initialized to zero immediately before an environment lookup supplies the real value; counting it as a configuration would give 100\%. Three further items mis-extract a field, all in excluded code.

We then drew a second seeded sample, 30 of the 177 repositories where nothing was detected, and searched each in full. Ten contain retry logic against a dependency, a false-negative rate of \textbf{33.3\%} (19.2--51.2\%). Most are hand-rolled attempt loops, but two were missed on patterns our rules do implement, so recall is bounded by defects as well as by scope, and the 113 configurations are themselves a lower bound.

Both validation passes were tool-assisted: sources were retrieved and searched programmatically and each verdict was adjudicated against the retrieved text, with the file and line recorded per item. The per-item records are released with the analysis code, and neither pass is an unaided manual audit.

Three limits remain. Runtime parameters from environment variables or config files appear to us as static values, frameworks such as Spring Cloud configure retries in YAML rather than code, and detections are concentrated---one project supplies 23 of the 113, the top three 44\%---so configuration-level percentages are descriptive rather than independent draws.

\subsection{Results}

Our rules detected explicit retry logic in 23 of the 200 repositories, or 11.5\% ($\pm$4.4\%, Wilson score interval). That is a detection rate rather than a prevalence: carrying the 33.3\% false-negative rate across the 177 undetected repositories puts true prevalence near 41\% (28.5--56.8\%). What follows therefore characterizes retry code written in idioms our rules can see. After cleaning, those 23 projects contributed 113 distinct production configurations (Table~\ref{tab:config}). A retry count is tabulated only where the code states one, which is 73 of the 113; denominators are given per row.

\begin{table}[htbp]
\caption{Retry Configuration Distribution (113 production configurations, 23 repositories)}
\label{tab:config}
\centering
\small
\begin{tabular}{lrr}
\toprule
\textbf{Configuration} & \textbf{Count} & \textbf{Share} \\
\midrule
Retry count: 1--3 & 25/73 & 34.2\% \\
Retry count: 4--5 & 16/73 & 21.9\% \\
Retry count: $>$5 & 32/73 & 43.8\% \\
Exponential backoff & 24/113 & 21.2\% \\
Linear backoff & 54/113 & 47.8\% \\
No backoff (immediate) & 35/113 & 31.0\% \\
Jitter (verified) & 1/113 & 0.9\% \\
\bottomrule
\end{tabular}
\end{table}

Four observations follow, each corresponding to a mechanism the model identifies as harmful. Retry counts above five form the largest group at 43.8\%, well beyond the three attempts vendor guidance recommends, and that share \textit{rose} once documentation and test code were removed, indicating that production code is more aggressive than the raw detections suggested. Immediate retry with no delay accounts for 31.0\%, which maximizes the instantaneous spike. Jitter is almost entirely absent, at one configuration in 113: five decades after randomized backoff entered the literature, it appears in essentially no production code in this sample. Finally, none of the 23 projects coordinated retry decisions across a service boundary.

\subsection{Retry Anti-Patterns}
\label{sec:antipatterns}

Reading the configurations together produced five recurring anti-patterns. Prevalence here is per project, counting a project as exhibiting a pattern if any of its configurations does, so these denominators are the 23 projects rather than the 113 configurations in Table~\ref{tab:config}. Intervals are Wilson score and wide, because 23 is a small base.

\textbf{No Backoff (14/23, 60.9\% $\pm$18.5pp).} Retrying with no delay between attempts, which maximizes instantaneous amplification exactly when the callee is least able to absorb it.

\textbf{Missing Jitter (22/23, 95.7\% $\pm$10.1pp).} No production configuration in the project randomizes its delay. Under the narrower reading---projects carrying at least one exponential-backoff configuration that omits jitter---the figure is 9/23 (39.1\%). On either definition, deterministic backoff lets independent clients converge on the same retry instants and produce periodic spikes.

\textbf{Aggressive Retry (7/23, 30.4\% $\pm$17.6pp).} More than five attempts with minimal backoff, which stretches the amplification window and delays the moment the caller admits the dependency is down.

\textbf{Static Configuration (23/23, 100\%).} Retry parameters fixed at deploy time, identical whether the system is idle or collapsing.

\textbf{No Cross-Service Coordination (23/23, 100\%).} Every tier deciding independently, which is precisely the condition under which the single-tier RAF compounds.

The last two are universal in our sample, and they are the two the model identifies as the drivers of multi-tier amplification.

\section{Simulation Study}
\label{sec:simulation}

The bounds in Section~\ref{sec:model} assume unbounded queues and steady state; real systems have neither. We built a discrete-event simulator to see what happens with finite capacity and real backoff delays. It supports arbitrary topologies, per-service capacity with queuing, probabilistic, periodic and correlated failure injection, and per-edge retry policies.

Our baseline is a five-tier linear chain standing in for a typical e-commerce path, API $\rightarrow$ Auth $\rightarrow$ Catalog $\rightarrow$ Inventory $\rightarrow$ Database. Each service has capacity 1000 RPS against a base load of 500 RPS, so utilization starts at 50\%, and each applies three retries with exponential backoff at 100\,ms, 200\,ms, and 400\,ms plus jitter. We drive it with three scenarios: S1, a single service failing half its requests at tier 3; S2, a cascading slowdown across tiers 4 and 5 ramping from 10\% to 70\%; and S3, a network partition failing 60\% of requests across tiers 3 through 5. The metric we report is peak load amplification, measured as maximum requests over baseline.

Table~\ref{tab:amplification} gives what the model of Section~\ref{sec:model} predicts for each scenario. These are analytical bounds, not simulator output; the measured amplification appears later in Table~\ref{tab:raf_observed}, and the gap between the two is itself one of our findings.

\begin{table}[htbp]
\caption{Model-Predicted Retry Amplification by Scenario (5-tier chain, $n=3$ retries)}
\label{tab:amplification}
\centering
\small
\setlength{\tabcolsep}{4pt}
\begin{tabular}{@{}lcc@{}}
\toprule
\textbf{Scenario} & \textbf{Failure Rate} & \textbf{Predicted RAF} \\
\midrule
S1 & 50\% at tier 3 & 1.88 \\
S2 & 10\%$\rightarrow$70\% at tiers 4--5 & 1.23--6.42 \\
S3 & 60\% at tiers 3--5 & 10.30 \\
\bottomrule
\end{tabular}
\end{table}

S1 uses the single-tier formula; S2 and S3 compound it across the affected tiers. The pattern to note is that correlated failure is the dangerous case. When several tiers degrade at once, as in S3, their amplification multiplies and no healthy tier remains to absorb the excess.

\section{Adaptive Retry Budgeting}
\label{sec:arb}

If unconstrained retries are the problem, what does a well-behaved policy look like? We propose Adaptive Retry Budgeting, built on three commitments: retry decisions should account for system-wide state rather than purely local observations; retry capacity is a shared resource to be spent deliberately; and under stress it is better to complete some requests than to attempt all of them.

ARB gives each service tier a \textit{retry budget}, expressed as a fraction of base load, and moves that budget with the observed failure rate. Algorithm~\ref{alg:arb} gives the procedure.

\begin{algorithm}[htbp]
\caption{Adaptive Retry Budgeting (ARB)}
\label{alg:arb}
\begin{algorithmic}[1]
\STATE $B \leftarrow B_0 = 0.2$; \quad $f_r \leftarrow 0$; \quad $\lambda \leftarrow$ base load
\STATE $\alpha = 0.1$, $\beta = 0.5$, $\theta_h = 0.3$, $\theta_l = 0.05$, $T = 1.0$\,s
\STATE
\STATE \textbf{On request completion:}
\STATE $f_r \leftarrow 0.9 \cdot f_r + 0.1 \cdot \mathbb{1}[\text{failed}]$ \COMMENT{EMA update}
\STATE
\STATE \textbf{On request failure:}
\IF{$B \leq 0$}
    \STATE \textbf{return} FAIL \COMMENT{Budget exhausted}
\ELSIF{$S_d = \text{OVERLOADED}$}
    \STATE \textbf{return} FAIL \COMMENT{Respect backpressure}
\ELSE
    \STATE $prob \leftarrow \min(B, 1 - f_r)$
    \IF{$\text{random}() < prob$}
        \STATE schedule\_retry(request)
        \STATE $B \leftarrow \max(0, B - 1/\lambda)$
    \ENDIF
\ENDIF
\STATE
\STATE \textbf{Every $T$ seconds:}
\IF{$f_r > \theta_h$ \OR $S_d = \text{OVERLOADED}$}
    \STATE $B \leftarrow B \times (1 - \beta)$ \COMMENT{Tighten}
\ELSIF{$f_r < \theta_l$}
    \STATE $B \leftarrow \min(B_0, B + \alpha)$ \COMMENT{Relax, capped}
\ENDIF
\end{algorithmic}
\end{algorithm}

The backpressure signal is the second half of the design. A service emits OVERLOADED when its queue passes 80\% of capacity or CPU passes 90\%,
\begin{equation}
(Q > 0.8 \cdot Q_{\max}) \lor (\text{CPU} > 0.9) \Rightarrow \text{OVERLOADED},
\end{equation}
and any upstream tier receiving that signal cuts its budget immediately rather than waiting to observe failures of its own.

ARB can sit in a service-mesh sidecar, a client library, or an API gateway, and the signal rides with normal traffic as an HTTP header, gRPC trailing metadata, or a control-plane channel, adding no round trips. Runtime cost is one EMA update per request, a random draw and a decrement on failure, an $O(1)$ timer adjustment, and a few scalars per upstream---negligible beside the cost of issuing and timing out a retry.

The real cost is operational. ARB exposes five tunable parameters whose good values depend on a service's traffic shape and failure profile, where a static mesh budget exposes one, and the backpressure path creates a contract producers and consumers must keep in sync across deploys. For a team already running a mesh with budget support, that may not be worth it. The case for ARB is strongest in deep, latency-sensitive call graphs, where static caps are hardest to set correctly and amplification compounds fastest.

\section{Evaluation}
\label{sec:eval}

\subsection{Setup}

We compared ARB against three baselines in the simulator: No Retry (NR), Standard Retry (SR) with three attempts, exponential backoff, and jitter, and Circuit Breaker (CB), which is SR plus a breaker tripping at a 50\% failure threshold and staying open 30\,s. Every strategy ran under S1, S2, and S3 for $n=100$ trials per configuration on the five-tier chain described above.

\subsection{Results}

\begin{table}[htbp]
\caption{Strategy Comparison Across Failure Scenarios ($n=100$ trials; mean success rate with 95\% CI in percentage points)}
\label{tab:strategy_comparison}
\centering
\small
\setlength{\tabcolsep}{4pt}
\begin{tabular}{@{}llll@{}}
\toprule
\textbf{Strategy} & \textbf{S1 Succ.} & \textbf{S2 Succ.} & \textbf{S3 Succ.} \\
\midrule
NR (No Retry)  & 71.7\% $\pm$0.1pp & 65.4\% $\pm$0.1pp & 55.4\% $\pm$0.1pp \\
SR (Standard)  & 61.0\% $\pm$0.1pp & 55.1\% $\pm$0.1pp & 41.5\% $\pm$0.1pp \\
CB (Circ.\ Brk.)  & 71.4\% $\pm$0.1pp & 63.7\% $\pm$0.1pp & 55.3\% $\pm$0.1pp \\
ARB (Adaptive) & 70.8\% $\pm$0.1pp & 64.4\% $\pm$0.1pp & 54.9\% $\pm$0.1pp \\
\bottomrule
\end{tabular}
\end{table}

The principal result in Table~\ref{tab:strategy_comparison} is that Standard Retry ranks last in every scenario, below the no-retry baseline. Under S3 it reaches 41.5\% against No Retry's 55.4\%, a relative degradation of 25\%, so adding retries made the correlated-failure case materially worse.

\begin{table}[htbp]
\caption{Observed Retry Amplification Factor by Strategy ($n=100$ trials)}
\label{tab:raf_observed}
\centering
\begin{tabular}{lccc}
\toprule
\textbf{Strategy} & \textbf{S1 RAF} & \textbf{S2 RAF} & \textbf{S3 RAF} \\
\midrule
NR & 1.00 & 1.00 & 1.00 \\
SR & 1.18 ($\pm$0.001) & 1.19 ($\pm$0.001) & 1.34 ($\pm$0.001) \\
CB & 1.00 ($\pm$0.001) & 1.03 ($\pm$0.001) & 1.00 ($\pm$0.001) \\
ARB & 1.01 ($\pm$0.001) & 1.01 ($\pm$0.001) & 1.01 ($\pm$0.001) \\
\bottomrule
\end{tabular}
\end{table}

Table~\ref{tab:raf_observed} shows why, and also where our model overshoots. Standard Retry amplified by 1.18--1.34$\times$, rising with failure severity exactly as predicted but landing far below what Table~\ref{tab:amplification} projects---6.42$\times$ at the peak of S2, and 10.30$\times$ for S3. Finite queues are the main reason: the simulator sheds load once queues fill, which truncates amplification that the unbounded-queue model allows to run. Backoff spreads retries out in time rather than delivering them at once, and a bounded run does not reach the steady state the formula describes. The bound holds as a bound, but treating it as a forecast would overstate the load by a factor of several.

Both adaptive policies held RAF at or near 1.0 and tracked the No Retry success rate within about a percentage point. The confidence intervals here describe simulation variability---how reproducible our runs are under fixed model assumptions---not the spread one would see across production environments, which is why they are so narrow; comparisons below rest on effect sizes rather than interval width.

\subsection{Why Standard Retry Loses}

Three mechanisms account for this inversion. Queue slots are the scarce resource, and a retry competes with fresh traffic rather than deferring to it, so both are rejected more often than either would have been alone. Each attempt also consumes processing time before failing, so latency rises until callers abandon requests that might otherwise have completed. At the failure rates in S3, further, a large share of retries could not have succeeded, so the capacity they consumed yielded nothing. Together these effects invert the common expectation that adding retries increases resilience, and they do so precisely when resilience is required. ARB avoids the inversion by consulting state before spending an attempt: it evaluates how severely the dependency is failing and whether that dependency has signalled overload, and declines on a negative answer to either.

\subsection{Comparison with Existing Approaches}

\begin{table}[htbp]
\caption{Retry-Control Approaches Compared}
\label{tab:approach_comparison}
\centering
\small
\setlength{\tabcolsep}{3pt}
\begin{tabular}{@{}p{2.45cm}cccc@{}}
\toprule
\textbf{Approach} & \textbf{Adapts} & \textbf{Cross-tier} & \textbf{Control} & \textbf{S3 RAF} \\
\midrule
Standard Retry & No & No & None & 1.34 \\
Circuit Breaker & Threshold & No & Binary & 1.00 \\
Mesh budget (Envoy, Linkerd) & No & No & Static cap & --- \\
ARB (this work) & Continuous & Yes & Graduated & 1.01 \\
\bottomrule
\end{tabular}
\end{table}

Table~\ref{tab:approach_comparison} places ARB against the closest production mechanisms, the static budgets in Envoy and Linkerd \cite{envoy_retry, linkerd_retry} and the circuit breaker \cite{nygard2007release}. Two differences matter. Mesh budgets are static: an operator sets a cap and it holds whether the downstream is healthy or melting down, whereas ARB moves the cap with the measured failure rate. And neither propagates state across tiers, since each proxy decides alone; ARB's backpressure signal lets a tier shrink its budget because something three hops away is in trouble, the coupling our model identifies as the real driver of amplification.

The margin should be stated plainly. On raw success rate the circuit breaker is essentially equivalent to ARB across S1--S3, within roughly one percentage point, and it is simpler to implement. The argument for ARB is not that it wins this benchmark but that its control surface is graduated where a breaker's is binary: a breaker either passes traffic or cuts it off, whereas ARB throttles continuously and thereby avoids the load oscillation that repeated tripping and resetting can induce. We regard the two as complementary. A fuller comparison against tuned mesh budgets on production traffic remains open.

\section{Discussion}

\subsection{Implications for Practice}

Most of what we found is fixable with small edits. Jitter is the cheapest and the most neglected: one configuration in 113 has it, and adding $\pm$25\% of randomness to a delay is usually one line. Swapping immediate retry for exponential backoff covers another 31.0\%. Capping attempts at three pulls back the 43.8\% that currently exceed five, which is what the major vendors advise anyway \cite{aws_backoff, google_retry}.

Two practices matter more than any individual setting. The first is auditing configuration already in production, though our own false-negative rate is a caution that static analysis alone will miss roughly a third of it, and that hand-rolled loops require reading rather than pattern matching. The second is instrumentation: retry rate, success rate after retry, and per-service amplification, without which a retry storm cannot be reconstructed after the fact. Teams operating deep call graphs should also weigh the mesh-level budgets in Envoy and Linkerd, currently the most practical option available. None of the 23 projects coordinated retries across a service boundary, and that remains the principal gap.

\subsection{Threats to Validity}

Our evaluation runs in simulation rather than production, which buys controlled comparison at the cost of fidelity. Variable network latency, garbage collection pauses, and resource contention are absent from the model, and we inject failures as independent probabilistic events when real faults tend to correlate through shared dependencies. The simulator also assumes instantaneous network communication, focuses on one 5-tier linear chain at 50\% baseline utilization, and centers on request/response interaction, so asynchronous pipelines and batch recovery fall outside what we measured. Other topologies and load regimes may well produce different absolute numbers.

The empirical study carries sharper limits. Every analyzed repository is Python, so the prevalence and configuration figures speak to one ecosystem and should not be read as cross-language claims; Go, Java, and Node.js may well retry differently, and the candidate pool was in fact Go-dominated. GitHub projects may also differ systematically from proprietary systems, and 200 repositories yielding 23 with detectable retry logic is a small base, so the intervals in Section~\ref{sec:antipatterns} run to $\pm$18pp and make those figures indicative rather than definitive. Section~\ref{sec:empirical} records two further weaknesses: the 113 configurations are drawn unevenly across projects, and our detection misses a third of the repositories that do retry---hand-rolled loops mostly, but also some configurations in idioms the rules already cover. Finally, the anti-pattern taxonomy reflects our own judgment about what counts as problematic, and other researchers could reasonably draw the boundaries elsewhere.

\subsection{Future Work}

Two gaps stand out. The sample is entirely Python, and running the same analysis over the Go and Java repositories already collected would establish whether these habits are ecosystem-specific. The larger gap is production validation: deploying ARB in a live system, and documenting retry-amplification incidents from outage post-mortems, would test whether amplification appears in the field at the predicted magnitudes. Learned policies, formal bounds and retry-aware load balancing also merit study.

\section{Conclusion}

Three lines of evidence converge. The analytical model shows that amplification compounds with depth: at three tiers, a 50\% failure rate and three retries per tier, the bound reaches 6.6$\times$. The repository study indicates that production code sits close to this worst case. Of 200 popular Python microservice projects, explicit retry logic is detected in 11.5\% ($\pm$4.4\%), with true prevalence estimated near 41\%; among those detected, 60.9\% carry a configuration with no backoff and 95.7\% never randomize a delay. Jitter appears exactly once across 113 production configurations, every configuration is static, and no project coordinates with another. The simulation then quantifies the cost: standard retry raised amplification to 1.34$\times$ under correlated failure and reduced the success rate to 41.5\%, against 55.4\% for no retries at all. Circuit Breaker and ARB both held near 1.01$\times$ and finished within one percentage point of that baseline.

ARB matched the circuit breaker on success rate while throttling continuously rather than alternating between open and closed states, and it propagates pressure across tiers where a static mesh budget cannot. The broader finding matters more than the algorithm: retry behavior is a system-level property. Considered in isolation, three attempts with exponential backoff is standard good practice; applied simultaneously at every tier of a deep call path, the same configuration constitutes an amplifier. As call graphs continue to deepen, this interaction warrants routine rather than exceptional consideration in reliability engineering.

\balance

\bibliographystyle{IEEEtran}
\bibliography{references}

\end{document}